\documentclass[aip,reprint]{revtex4-1}

\usepackage{graphicx}
\usepackage{dcolumn}
\usepackage{bm}

\draft 

\begin{document}

\title{A numerical ellipsometric analysis (NEA) of nanoscale layered systems} 

\author{Giuseppe Emanuele Lio, Giovanna Palermo, Roberto Caputo and Antonio De Luca}
\email[]{E-mail: giuseppe.lio@unical.it}
\affiliation{CNR-Nanotec, Cosenza and Physics Department, University of Calabria, \\87036 Arcavacata di Rende (CS), Italy}


\begin{abstract}
A simple and robust method able to predict, with high accuracy, the optical properties of single and multi-layer nanostructures is presented. The method exploits a COMSOL Multiphysics simulation platform and it has been validated by three case studies with increasing numerical complexity: i) a single thin layer (20 nm) of Ag deposited on a glass substrate; ii) a metamaterial composed of five bi-layers of Ag/ITO (Indium Tin Oxide), with a thickness of 20 nm each; iii) a system based on a three-materials unit cell (AZO/ITO/Ag), but without any thickness periodicity (AZO stands for Al$_2$O$_3$/Zinc Oxide). Numerical results have been compared with experimental data provided by real ellipsometric measurements performed on the above mentioned nanostructures ad-hoc fabricated. The obtained agreement is excellent suggesting this research as a valid approach to design materials able to work in a broad spectrum range.
\end{abstract}


\maketitle 

\section{Introduction}
In the last years, the complexity of systems focus of materials science and experimental physics is gradually increasing to the point that an analytical approach to the theoretical modeling is always more leaving the field to its numerical counterpart. This trend in research is thus promoting the evolution of comprehensive numerical packages that provide an extremely accurate description of physical processes. 
Within the portfolio of experimental characterization techniques, it is remarkable what a detailed insight can be acquired when performing a spectroscopic ellipsometry of thin films and bulk materials, and determination of general optical material characteristics. Indeed, after design and realization of a specific system, an ellipsometric characterization typically represents a crucial testbed providing a validation of the expected functionalities or helpful hints for improvement. At this stage, a trial-and-error procedure to achieve the desired target very often results in numerous attempts. Considering the vast range of applicability of ellipsometry, the possibility of implementing this technique as a reliable numerical method, able to a priori predict the result of the real experimental analysis before performing it, evidently results extremely convenient. This is especially true when the focus of the study is a complex nanostructured system whose fabrication can be largely time-consuming and expensive.
Here, we present a novel method, implemented in the COMSOL Multiphysics, providing a comprehensive ellipsometric analysis of general multi-layer systems with extreme freedom of design in terms of thickness, composition and number of layers. \\
For each considered system, the proposed numerical ellipsometric analysis (NEA)\cite{Lio2019cavities} allows calculating the main optical quantities of interest for different polarization and angle of incidence. 

\section{Results and Discussion} 

In order to validate the effectiveness of the NEA modeling, numerical predictions have been compared with results of measurements performed on the corresponding real samples. Several multi-layer systems have been designed, fabricated and characterized by an experimental ellipsometric analysis. 

\begin{figure}[h]
\begin{center}
\includegraphics[width=1\columnwidth]{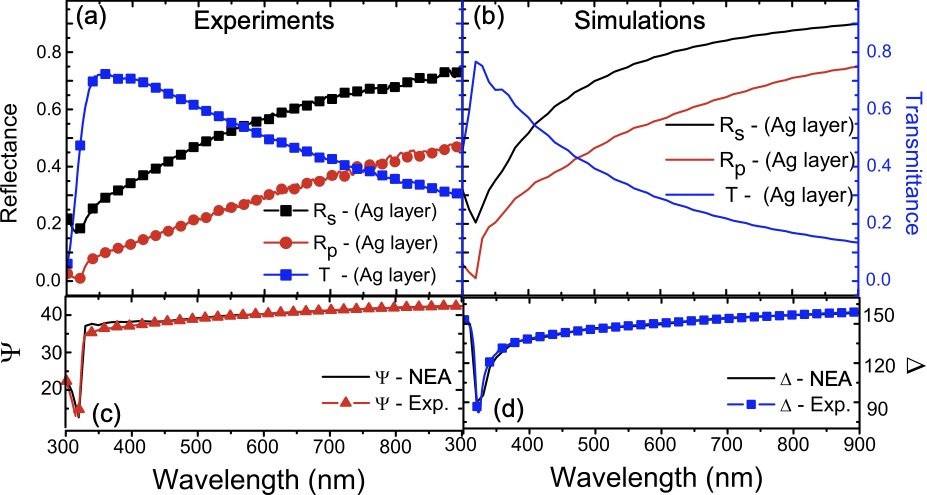}
  \caption{ Reflectances ($R_p$, $R_s$, red and black lines with symbols) and transmittance ($T$ blue lines with squares) measured (a) and calculated by the NEA model (b) for a single Ag layer on a glass substrate. Comparison between $\Psi$ and $\Delta$ for the experimental (colored triangles and squares) and numerical (black lines) case respectively, referred to the same sample (c-d). Figure readapted from ref [1]}
  \label{2}
  \end{center}
\end{figure}

The first and simplest system is a single Ag layer deposited on a glass substrate (20nm thickness). The following system is a hyperbolic metamaterial made of a stack of five alternated Ag/ITO bi-layers (ITO stands for Indium Tin Oxide) with the single bilayer having a thickness of 40nm (20/20). The third system is characterized by a three-materials unit cell repeated three times. The unit cell is made of Al$_2$O$_3$ doped Zinc Oxide (AZO), ITO and Ag. In order to get the numerical task more difficult, each of the three cells constituting the system has different thickness for each layer. All samples have been experimentally fabricated by exploiting a DC sputtering technique. Then, for all cases the experimental reflectance, transmittance and ellipsometric angles $\Psi$ and $\Delta$ have been measured by means of a M-2000 ellipsometer (J.A. Woollam).
Figures \ref{2}a and \ref{2}b respectively show the directly measured reflectance and transmittance, as well as the corresponding numerical curves provided by the NEA simulation for the first case study. The reflectances, $R_p$ and $R_s$ curves, are measured and calculated by considering an incident angle $\theta_{i}$ = $50^\circ$, while the transmittance was measured and simulated at normal incidence. The minimum value at about 330nm in the reflectance curves (and the related maximum in the transmittance ones) is referred to as the Ferrel-Berreman mode for a silver nanometric layer.

\begin{figure}[h]
\begin{center}
\includegraphics[width=1\columnwidth]{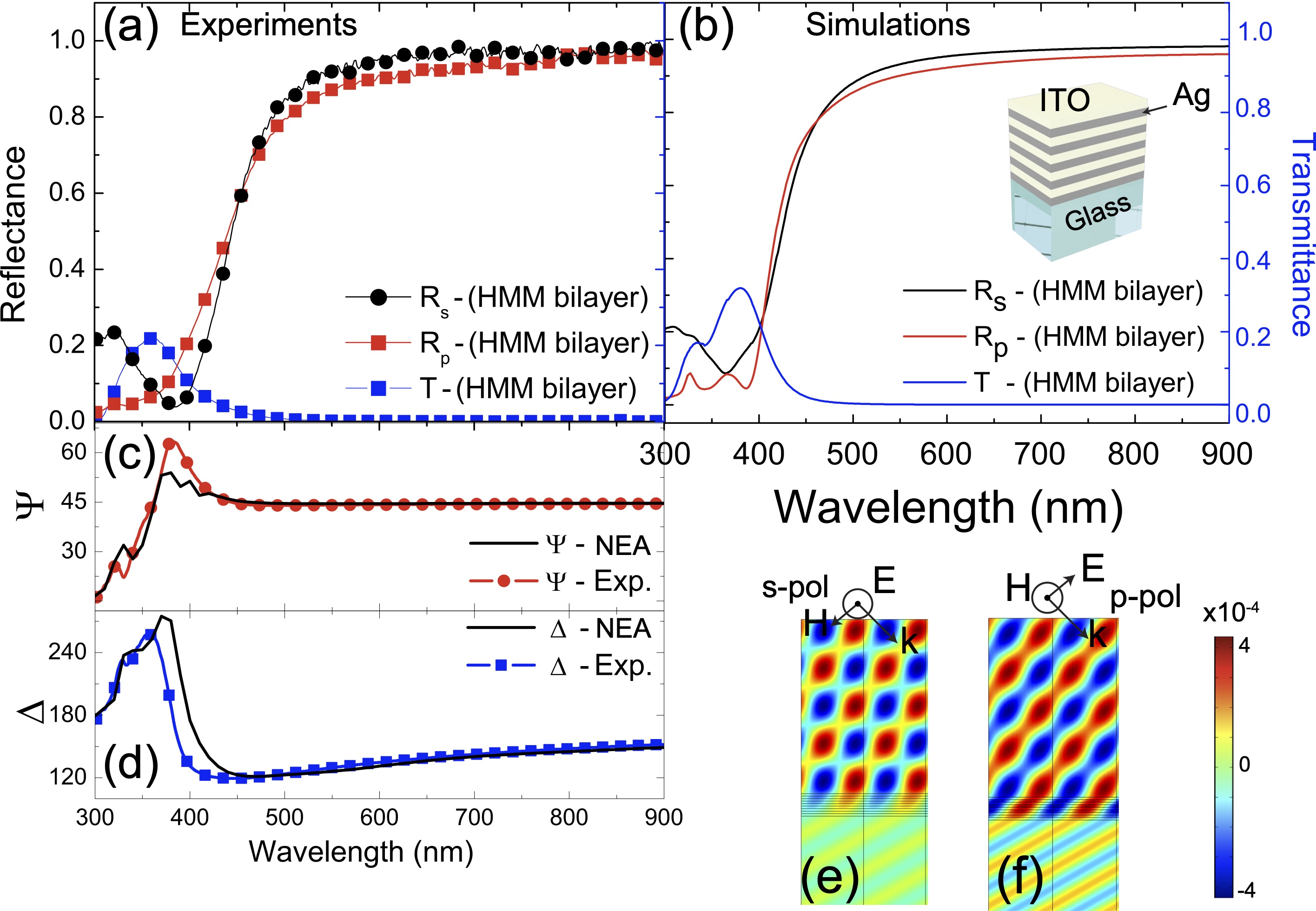}
  \caption{Reflectances ($R_p$, $R_s$ red and black lines with symbols) and transmittance ($T$ blue lines with squares) measured (a) and calculated by the NEA model (b) for a five bilayers Ag/ITO HMM, sketched in the inset. Comparison between experimental (colored lines with symbols) and numerical (solid lines) $\Psi$ (c) and $\Delta$ (d), for the same sample. (e-f) Electric field maps distribution for TE (s-pol) and TM (p-pol) waves respectively, extracted for an impinging wavelength of 390nm.Figure readapted from ref [1]}
  \label{3}
  \end{center}
\end{figure}

\begin{figure}[h]
\begin{center}
\includegraphics[width=1\columnwidth]{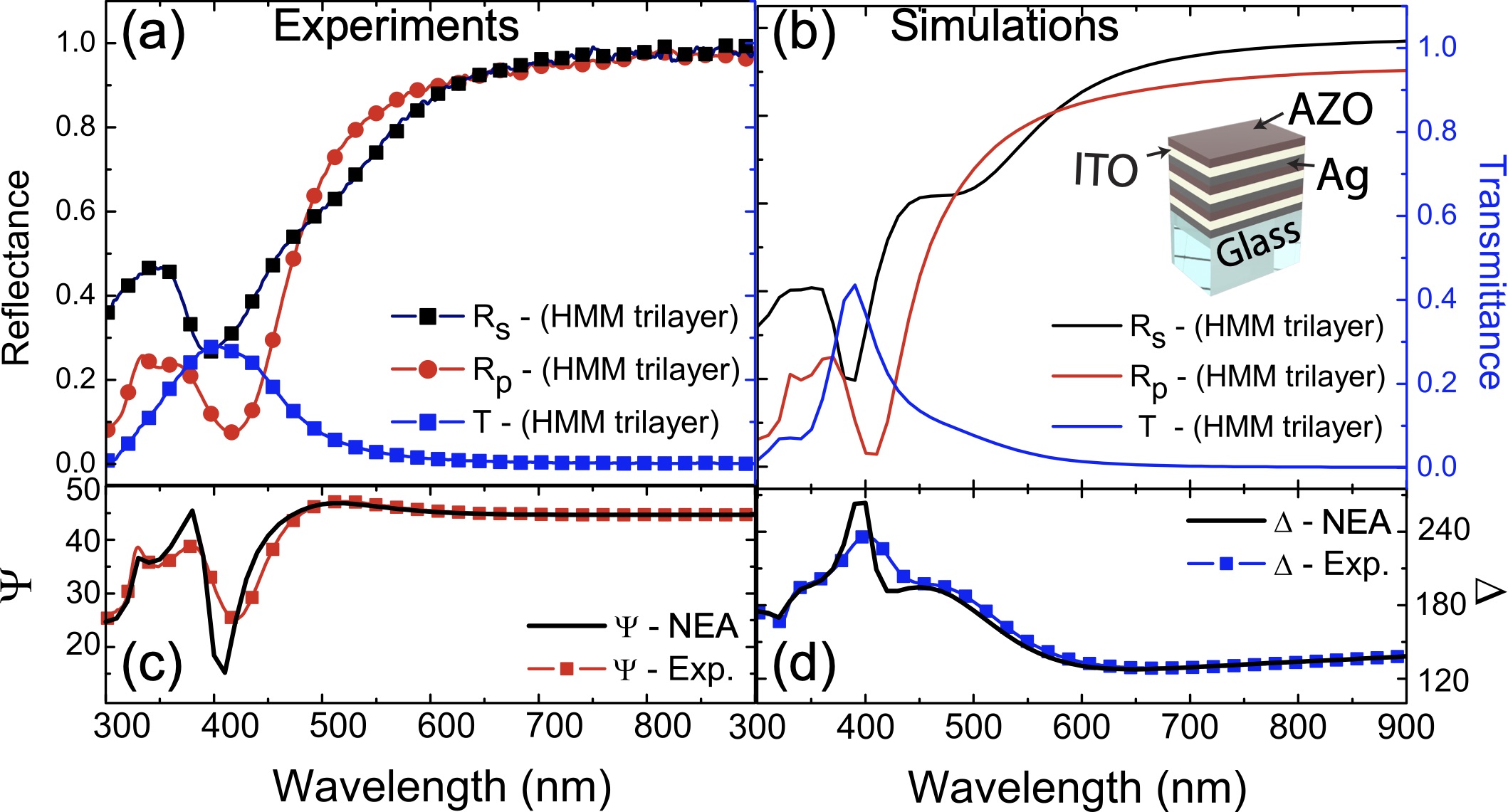}
\caption{Reflectances ($R_p$, $R_s$, red and black lines with symbols) and transmittance ($T$ blue lines with squares) measured (a) and calculated by the NEA model (b) for the metamaterial with a three-materials unit cell (AZO/ITO/Ag). In the inset a 3D sketch of the sample is reported. Comparison between experimental (colored lines with symbols) and numerical (solid lines) $\Psi$ (c) and $\Delta$ (d), for the same sample.Figure readapted from ref [1]}
\label{4}
\end{center}
\end{figure}

The difference of about $20\%$ between the amplitude of experimental and simulated curves is due to the presence of the glass substrate in the real sample, partially considered in the simulated case. Indeed, in order to optimize the computational time the size of the glass in the simulation was decreased to $1200nm$ instead of the real $1.1mm$. The result only differs in absolute values but not in the general trend of the curves. Apart from that, the agreement between measurements and simulations is quite satisfactory.
Figures \ref{2}c and \ref{2}d show the $\Psi$ and $\Delta$ behavior, respectively. 
The second test of our NEA model has been conducted on a hyperbolic metamaterial (HMM) that, by alternating lossy metal layers to dielectric ones, acquires particular optical features. Thickness and composition of the nanolayers can be opportunely designed in order to exploit unusual optical properties in a desired spectral range. In fact, the particular design and choice of sizes and constitutive materials leads to the \textit{epsilon-near-zero-and-pole} ($\varepsilon_{NZP}$) HMM, allowing extraordinary light confinement properties. The proposed HMM system is sketched in the inset of Figure \ref{3}b and is composed of five Ag/ITO bilayers, characterized by a fill fraction of 50\%.
Figures \ref{3}a  and \ref{3}b show, respectively, the experimental and numerical reflectances ($R_p$, $R_s$) and transmittance ($T$) calculated by the NEA model together with the comparison between experimental and numerical $\Psi$ and $\Delta$ for the same sample (Figures \ref{3}c-\ref{3}d). Also in this case, all measured quantities find an impressive agreement with the numerical counterpart.
Figures \ref{3}e and \ref{3}f also report the electric field maps respectively for an s-pol and a p-pol wave at $\lambda=390nm$, where it is well evident how the p-pol wave is able to penetrate more efficiently through the HMM structure than in the s-pol case, being transmitted by the medium in accordance with the metamaterial prediction.
A further and more complex test of the NEA model has been performed by considering the third system where every kind of thickness periodicity is absent. This choice reflects a more typical experimental situation with different thicknesses of all involved layers and materials. The sample is composed of a unit cell of three materials, Ag, ITO and AZO, repeated three times, whereas the thickness of each layer is different (see inset of Figure \ref{4}b). Starting from the glass substrate, the layers thicknesses are: 19 nm (Ag), 15 nm (ITO), 15 nm (AZO), 20 nm (Ag), 21 nm (ITO), 16 nm (AZO), 23 nm (Ag), 27 nm (ITO) and 15 nm (AZO).
Figure \ref{4} reports the results related to this sample; the agreement between experiments and simulations is quite remarkable also in this case.

\section{Conclusions}
In this contribution, we propose a solid numerical tool exploiting the COMSOL Multiphysics platform able to predict the optical behavior of real case studies. The tool has been validated in presence of systems with increasing structural complexity, ranging from single metal layers to aperiodic multi-layers. The resulting comprehensive optical and ellipsometric analysis of considered nanoscale structures has shown an excellent agreement between numerical and experimental curves depicting the main optical features as reflectance, transmittance, as well as the ellipsometric angles $\Psi$ and $\Delta$. This confirms the effectiveness of the tool as a significant instrument for nanotechnology design and fundamental research in nanoscience. 

\section*{Methods: The numerical ellipsometer}
\begin{figure}[h]
\begin{center}
\includegraphics[width=0.8\columnwidth]{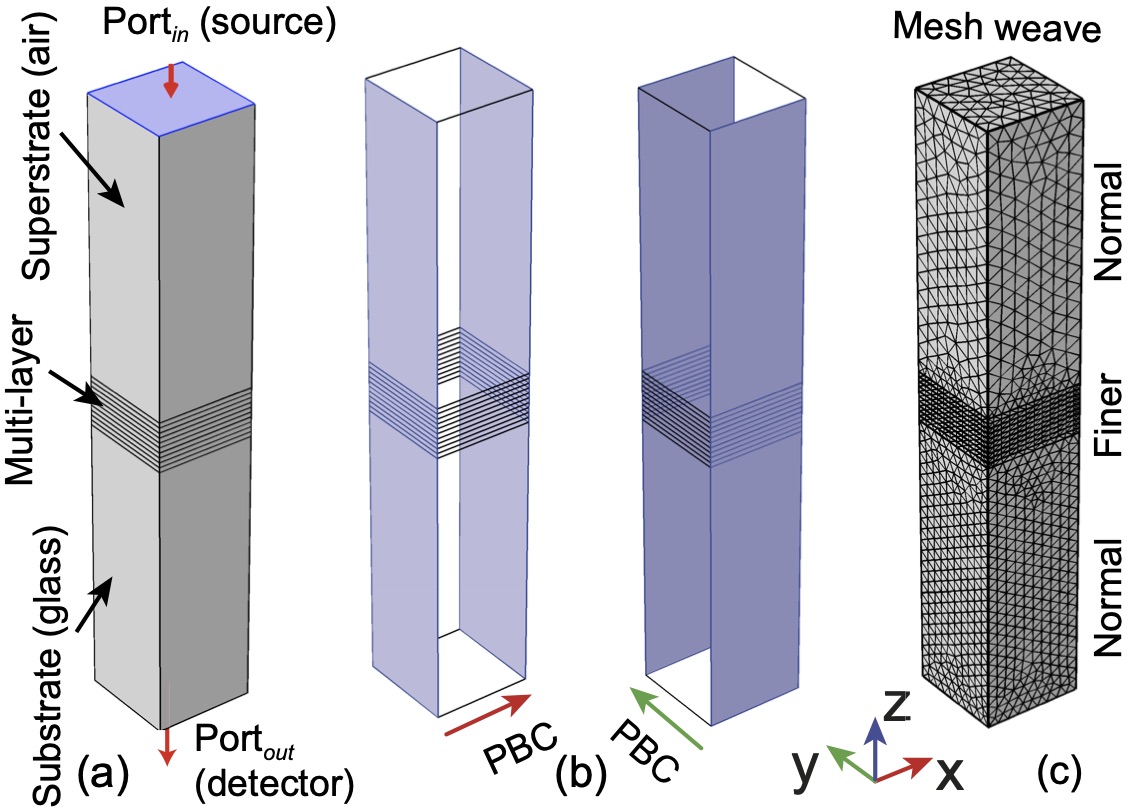}
  \caption{a) The main structure of the NEA model, it is realized from a parallelepiped divided in layers that constitute the superstrate, the layer of materials and the substrate. (b) Here it is shown the application direction of the periodic boundary condition from periodic port. (c) A sketch of the mesh directly controlled from the physics, the larger part are meshed with a normal weave while for the small parts has been used a fine mesh weave. Figure readapted from ref [1]}
  \label{1}
  \end{center}
\end{figure}

In order to implement the numerical ellipsometer analysis (NEA) of single or multi-layered systems in COMSOL, the geometry shown in Figure\ref{1}a is to be considered. An electromagnetic plane wave  impinges on a generic multi-layer system from the superstrate (typically air), with a specific incident angle and polarization state (s-polarization (TE) or p-polarization (TM)). Light interacting with the structure is then collected through both its substrate (typically glass) and superstrate, permitting to compute the optical quantities of interest (transmittance (T), reflectance (R), Brewster angle, field maps and so on). This is done by solving the frequency-domain partial differential equation (PDE) that governs the $\mathbf{E}$ and $\mathbf{H}$ fields associated with the electromagnetic wave propagating through the structure.
Later on, from the field values, the relevant quantities can be derived. Indeed, in case of a standard ellipsometric analisys, we have to measure the ratio $\rho =\widetilde{r}_{p}/ \widetilde{r}_{s}$, where $\widetilde{r}_{p}$ and $\widetilde{r}_{s}$ are the complex Fresnel coefficients related to the square roots of the reflectances for the two different polarizations, $\sqrt{R_p}=r_p=Re (\widetilde{r}_{p})$ and $\sqrt{R_s}=r_s=Re (\widetilde{r}_{s})$. Being $\rho$ a complex quantity, it can be written also as $\tan(\Psi )e^{i\Delta }$. Thus, $\tan(\Psi)$ is related to the amplitude ratio upon reflection ($\Psi=arctan(r_p/r_s)$), whereas $\Delta$ represents the phase shift between the two components ($\Delta=\phi_{p} - \phi_{s}$).\cite{tompkins2005handbook, losurdo2013ellipsometry} Since ellipsometry is measuring the ratio (or difference) of two values (rather than their absolute values), it is very robust, accurate and reproducible. $\Delta$ is evaluated with the following equation:  $\Delta= [\Im(\ln{\widetilde{r}_{p} / \widetilde{r}_{s})+\pi}]$. 

In order to simulate the behavior of TE (s-pol) and TM (p-pol) polarized light in a 2D environment, it is necessary to properly write the components of the electromagnetic fields with respect to the incidence plane ($xy$). To select the s-polarization in COMSOL, an "out of plane" configuration has to be used $(\mathbf{E}=(0,0,1))$, while to select the p-polarization, the user has to set the "in plane wave-vector" scheme; in this case the magnetic field has to be used as $\mathbf{H}=(0,0,1)$. In a 3D simulation, the incident plane corresponds to the $xz$ plane and the polarizations are set accordingly: $\mathbf{E}=(0,1,0)$ for s-pol or $\mathbf{H}= (0,1,0)$ for p-pol.
To optimize the computation time of the NEA calculation, Floquet periodic boundary conditions (PBCs) have been set to all boundaries of the geometry ($x$ and $y$ directions), paired two-by-two (Figure \ref{1}b). Another way to optimize the computation process consists in parallelizing the solvers by introducing two "frequency domain studies", able to solve the electromagnetic equation for both s- and p-polarized light simultaneously. An important role in the numerical system is carried out by the mesh that discretizes the problem. COMSOL gives the opportunity to set it as controlled by the physics, by selecting the option "physics-controlled the mesh" in "physics study" menu. A "normal mesh" is automatically generated by the software, depending on the minimal size present in the system, compared with the incident wavelength Figure \ref{1}c.

\section*{Acknowledgements}
The authors thank the ``Area della Ricerca di Roma 2", Tor Vergata, for the access to the ICT Services (ARToV-CNR) for the use of the COMSOL Multiphysics Platform, Origin Lab and Matlab, and the Infrastructure ``BeyondNano" (PONa3-00362) of CNR-Nanotec for the access to research instruments.

\bibliography{Ellips_AIP} 

\begin{thebibliography}{3}%
\makeatletter
\providecommand \@ifxundefined [1]{%
 \@ifx{#1\undefined}
}%
\providecommand \@ifnum [1]{%
 \ifnum #1\expandafter \@firstoftwo
 \else \expandafter \@secondoftwo
 \fi
}%
\providecommand \@ifx [1]{%
 \ifx #1\expandafter \@firstoftwo
 \else \expandafter \@secondoftwo
 \fi
}%
\providecommand \natexlab [1]{#1}%
\providecommand \enquote  [1]{``#1''}%
\providecommand \bibnamefont  [1]{#1}%
\providecommand \bibfnamefont [1]{#1}%
\providecommand \citenamefont [1]{#1}%
\providecommand \href@noop [0]{\@secondoftwo}%
\providecommand \href [0]{\begingroup \@sanitize@url \@href}%
\providecommand \@href[1]{\@@startlink{#1}\@@href}%
\providecommand \@@href[1]{\endgroup#1\@@endlink}%
\providecommand \@sanitize@url [0]{\catcode `\\12\catcode `\$12\catcode
  `\&12\catcode `\#12\catcode `\^12\catcode `\_12\catcode `\%12\relax}%
\providecommand \@@startlink[1]{}%
\providecommand \@@endlink[0]{}%
\providecommand \url  [0]{\begingroup\@sanitize@url \@url }%
\providecommand \@url [1]{\endgroup\@href {#1}{\urlprefix }}%
\providecommand \urlprefix  [0]{URL }%
\providecommand \Eprint [0]{\href }%
\providecommand \doibase [0]{http://dx.doi.org/}%
\providecommand \selectlanguage [0]{\@gobble}%
\providecommand \bibinfo  [0]{\@secondoftwo}%
\providecommand \bibfield  [0]{\@secondoftwo}%
\providecommand \translation [1]{[#1]}%
\providecommand \BibitemOpen [0]{}%
\providecommand \bibitemStop [0]{}%
\providecommand \bibitemNoStop [0]{.\EOS\space}%
\providecommand \EOS [0]{\spacefactor3000\relax}%
\providecommand \BibitemShut  [1]{\csname bibitem#1\endcsname}%
\let\auto@bib@innerbib\@empty
\bibitem [{\citenamefont {Lio}\ \emph {et~al.}(2019)\citenamefont {Lio},
  \citenamefont {Palermo}, \citenamefont {Caputo},\ and\ \citenamefont
  {De~Luca}}]{Lio2019cavities}%
  \BibitemOpen
  \bibfield  {author} {\bibinfo {author} {\bibfnamefont {G.~E.}\ \bibnamefont
  {Lio}}, \bibinfo {author} {\bibfnamefont {G.}~\bibnamefont {Palermo}},
  \bibinfo {author} {\bibfnamefont {R.}~\bibnamefont {Caputo}}, \ and\ \bibinfo
  {author} {\bibfnamefont {A.}~\bibnamefont {De~Luca}},\ }\bibfield  {title}
  {\enquote {\bibinfo {title} {A comprehensive optical analysis of nanoscale
  structures: from thin films to asymmetric nanocavities},}\ }\href {\doibase
  10.1039/C9RA03684A} {\bibfield  {journal} {\bibinfo  {journal} {RSC
  Advances}\ }\textbf {\bibinfo {volume} {9}},\ \bibinfo {pages} {21429--21437}
  (\bibinfo {year} {2019})}\BibitemShut {NoStop}%
\bibitem [{\citenamefont {Tompkins}\ and\ \citenamefont
  {Irene}(2005)}]{tompkins2005handbook}%
  \BibitemOpen
  \bibfield  {author} {\bibinfo {author} {\bibfnamefont {H.}~\bibnamefont
  {Tompkins}}\ and\ \bibinfo {author} {\bibfnamefont {E.~A.}\ \bibnamefont
  {Irene}},\ }\href@noop {} {\emph {\bibinfo {title} {Handbook of
  ellipsometry}}}\ (\bibinfo  {publisher} {William Andrew},\ \bibinfo {year}
  {2005})\BibitemShut {NoStop}%
\bibitem [{\citenamefont {Losurdo}\ and\ \citenamefont
  {Hingerl}(2013)}]{losurdo2013ellipsometry}%
  \BibitemOpen
  \bibfield  {author} {\bibinfo {author} {\bibfnamefont {M.}~\bibnamefont
  {Losurdo}}\ and\ \bibinfo {author} {\bibfnamefont {K.}~\bibnamefont
  {Hingerl}},\ }\href@noop {} {\emph {\bibinfo {title} {Ellipsometry at the
  nanoscale}}}\ (\bibinfo  {publisher} {Springer},\ \bibinfo {year}
  {2013})\BibitemShut {NoStop}%
\end{thebibliography}%
\end{document}